\documentclass[runningheads]{svmult}

\usepackage{makeidx}   
\usepackage{graphicx}  
\usepackage{subeqnar}  
\usepackage{multicol}  
\usepackage{physprbb}  
\makeindex             

\newcommand{\Msolar}{\mbox{\,$\rm M_{\odot}$}} 
\begin{document}
\title*{On the black hole - bulge mass relation in active and inactive
galaxies}
\toctitle{On the black hole - bulge mass relation in active and inactive
galaxies}
%
%
\titlerunning{The black hole - bulge mass relation}
%
\author{R.J. McLure\inst{1}
\and J.S. Dunlop \inst{2}}
\authorrunning{R.J. McLure \& J.S. Dunlop}
%
%
\institute{Oxford University, UK
\and Institute for Astronomy,
     Edinburgh University,
     UK}

\maketitle              

\begin{abstract}
New virial black-hole mass estimates are presented for a
sample of 72 AGN covering three decades in optical luminosity.  
Using a model in which the AGN broad~-~line region (BLR) has a flattened
geometry, we investigate the $M_{bh}-$~$L_{bulge}$ relation for a combined 90-object sample,
consisting of the AGN plus a sample of 18 nearby inactive elliptical galaxies
with dynamical black-hole mass measurements. It is found that, for all
reasonable mass-to-light ratios, the $M_{bh}-L_{bulge}$ relation
is equivalent to a linear scaling between bulge and black-hole
mass. The best-fitting normalization of the $M_{bh}-M_{bulge}$
relation is found to be $M_{bh}=0.0012M_{bulge}$, in agreement with
recent black-hole mass studies based on stellar velocity dispersions.
Furthermore, the scatter around the $M_{bh}-L_{bulge}$
relation for the full sample is found to be significantly smaller than
has been previously reported ($\Delta \log M_{bh}=0.39$ dex). Finally,
using the nearby inactive elliptical galaxy sample alone, it is shown
that the scatter in the $M_{bh}-L_{bulge}$ relation is only $0.33$ dex,
comparable to that of the $M_{bh}-\sigma$ relation. These results
indicate that reliable black-hole mass estimates can be obtained for
high redshift galaxies.
\end{abstract}
\section{Introduction}
\label{intro}
The correlation between black-hole mass and bulge luminosity 
is now well established for both active and inactive galaxies 
(Magorrian et al. 1998; Laor 1998). However, despite recent 
attention in the literature, the usefulness of the $M_{bh}-L_{bulge}$ relation as a
black-hole mass estimator is at present severely limited due to its 
large scatter ($\simeq0.5$ dex). Although the correlation between 
black-hole mass and stellar velocity dispersion for nearby 
inactive galaxies displays a much smaller
scatter ($\simeq0.3$ dex, Merritt \& Ferrarese
2001a), it is clear that a $M_{bh}-L_{bulge}$ correlation with
reduced scatter would be highly desirable, given the extreme difficulty in
obtaining stellar velocity dispersions for high redshift galaxies. 

This conference proceeding presents the main results of a new study
(McLure \& Dunlop 2002) in which we investigate the black hole - bulge
mass relation using a 90-object sample comprised of 72 AGN (53 QSOs
and 19 Seyfert 1s) and 18 nearby quiescent ellipticals with
dynamically determined black-hole mass estimates. Those interested in
the details of our analysis, particularly the flattened geometry model
adopted for the calculation of the virial black-hole mass estimates, are
referred to McLure \& Dunlop (2002).
\section{The black-hole mass - bulge luminosity relation}
\begin{figure}[b]
\begin{center}
\includegraphics[width=.5\textwidth,angle=270]{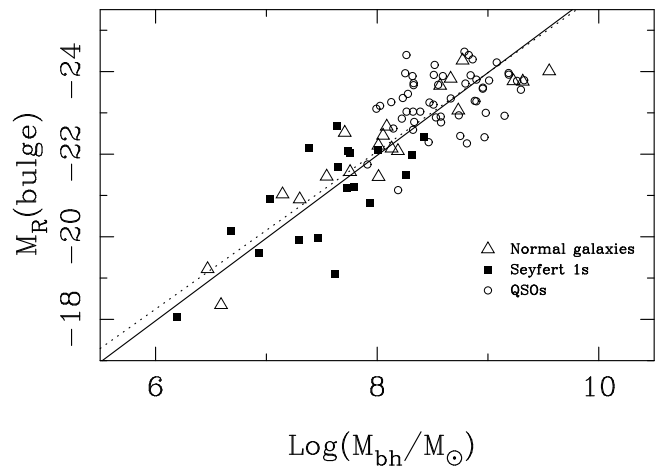}
\end{center}
\caption[]{Absolute $R-$band bulge magnitude versus black-hole mass for
the full 90-object sample. The black-hole masses for the 72 AGN are
derived from their H$\beta$ line-widths under a disc-like BLR
model (see McLure \& Dunlop 2002). The black-hole masses of the
inactive galaxies (triangles) are
dynamical estimates as compiled by Kormendy \& Gebhardt (2001). Also
shown is the formal best-fit (solid line) and the best-fitting
linear relation (dotted line).}
\label{fig1}
\end{figure}
In Fig \ref{fig1} absolute $R-$band bulge magnitude is plotted
against black-hole mass for the 72 objects in the AGN
sample. Also shown is absolute $R-$band bulge
magnitude plotted against dynamically-estimated black-hole mass for 
our nearby inactive elliptical galaxy sample. Several aspects of Fig
\ref{fig1} are worthy of immediate comment. Firstly, as was shown by
McLure \& Dunlop (2001) and by Laor (1998 \& 2001), it can be seen
that bulge luminosity and
black-hole mass are extremely well correlated, with 
$r_{s}=-0.77$ ($7.3\sigma$). Secondly, it
is clear that the AGN and nearby inactive galaxy samples follow the
same $M_{bh}-L_{bulge}$ relation over $>3$ decades in black-hole mass,
and $>2.5$ decades in bulge luminosity. This second fact strongly
supports the conclusions of Dunlop et
al. (2001) and Wisotzki et al. (2001), that the host-galaxies of 
powerful quasars are normal massive ellipticals drawn from the bright
end of the elliptical galaxy luminosity function. Thirdly, there
can be seen to be no systematic offset between the Seyfert 1 and 
quasar samples, reinforcing the finding of McLure \& Dunlop (2001) 
that, contrary to the results of Wandel (1999), the bulges of 
Seyfert galaxies and QSOs form a continuous
sequence which ranges from $M_{R}$(bulge)$\simeq-18$ to
$M_{R}$(bulge)$\simeq-24.5$. If we adopt an integrated value of 
$M_{R}^{\star}=-22.2$ (Lin et
al. 1998), then this implies that the $M_{bh}-L_{bulge}$
relation holds from $L_{bulge}\simeq 0.01L^{\star}$, all the way up
to objects which constitute some of the most massive ellipticals ever
formed; $L_{bulge}\simeq 10L^{\star}$. 

The best-fit to the full 90-object sample has the following form:
\begin{equation}
\log(M_{bh}/\Msolar)=-0.50(\pm0.02)M_{R}-2.96(\pm0.48)
\label{bestfit}
\end{equation}
\noindent
and is shown as the solid line in Fig \ref{fig1}. The scatter around
this best-fitting relation is only $\Delta M_{bh}=0.39$ dex, an uncertainty
factor of $<2.5$. The reduced scatter found here in comparison to
previous studies is due to two factors. Firstly, all of the bulge
luminosities used in this study are derived from full two-dimensional
modelling of high resolution data, the majority of which is from HST. 
The second factor is the inclination corrections to the black-hole
mass estimates provided by our flattened-geometry BLR model. Both of
these aspects are discussed in detail in McLure \& Dunlop (2002).

Given that the 18 objects in the nearby inactive
galaxy sample have actual dynamical black-hole mass estimates, it is 
obviously of interest to quantitatively test how consistent the 
$M_{bh}-L_{bulge}$ relation for these objects is with the fit to 
the full, AGN dominated, sample. The best-fit to the inactive 
galaxy sub-sample alone, has the following form:
\begin{equation}
\log(M_{bh}/\Msolar)=-0.50(\pm0.05){\rm M}_{\rm R}-2.91(\pm1.23)
\end{equation}
\noindent
which can be seen to be perfectly consistent with the best-fit to the
full sample in terms of both slope and normalization. Indeed, 
the best-fitting relations for the full sample,
quasar sample, Seyfert galaxy sample and the nearby inactive 
galaxy sample are all internally consistent, and display comparable
levels of scatter.  This is a remarkable result 
given that it implies that the combined bulge/black hole formation 
process was essentially the same throughout the full sample, which as well 
as featuring both active and inactive galaxies, includes 
galaxies of both late and early-type morphology.
\subsection{The linearity of the black hole - bulge mass relation}
In our previous study (McLure \& Dunlop 2001) of a sample of 45 AGN we found
that $M_{bh} \propto M_{bulge}^{1.16\pm0.16}$, and therefore
concluded that there was no evidence that the $M_{bh}-M_{bulge}$
relation was non-linear. In contrast, evidence for a non-linear relation was
recently found by Laor (2001). In his $V-$band study of the black hole to
bulge mass relation in a 40-object sample (15 PG quasars, 16 inactive
galaxies and 9 Seyfert galaxies) Laor found a best-fitting relation of
the form $M_{bh}=M_{bulge}^{1.54\pm0.15}$, which is clearly
inconsistent with linearity. However, 
in order to determine
the $M_{bh}-M_{bulge}$ relation it is obviously necessary to convert
the measured bulge luminosities into masses, via an adopted
mass-to-light ratio. The form of this mass-to-light ratio affects the derived
slope of the $M_{bh}-M_{bulge}$ relation in the following way. If
the mass-to-light ratio is parameterized as $M/L \propto L^{\alpha}$,
then the resulting slope ($\gamma$) of the $M_{bh}-M_{bulge}$
relation is given by $\gamma=\frac{-2.5 \beta}{1+\alpha}$, where
$\beta$ is the slope of the $M_{bh}-L_{bulge}$ relation (Eqn 1).

Here we choose to adopt the derived $R-$band mass-to-light ratio
for the Coma cluster from J$\o$rgensen, Franx \& Kj$\ae$rgaard (1996),
which has $\alpha=0.31$. With this mass-to-light ratio our 
best-fitting $M_{bh}-L_{bulge}$ relation 
transforms to a $M_{bh}-M_{bulge}$ relation of the following form:
\begin{equation}
M_{bh} \propto M_{bulge}^{0.95\pm0.05}
\end{equation}
\noindent
It can immediately be seen that from our results there is no
indication that the scaling between black hole and bulge
mass is non-linear. 

In order to calculate the bulge mass of the objects in his sample, Laor 
(2001) adopted a $V-$band mass-to-light ratio of 
$M_{bulge}\propto L_{bulge}^{1.18}$ (Magorrian et al. 1998), which 
is significantly different from our chosen mass-to-light
ratio. However, irrespective of this, our new best-fit to the slope of the 
$M_{bh}-L_{bulge}$ relation ($\beta=-0.50\pm0.02$) of our new sample,
which has a larger dynamic range in $L_{bulge}$ than both the samples
studied in McLure \& Dunlop (2001) and Laor (2001), means that any disagreement about
mass-to-light ratios cannot now alter the conclusion that the
$M_{bh}-M_{bulge}$ relation is consistent with being linear. To
demonstrate this we conclude by noting that even using 
the $M_{bulge}\propto L_{bulge}^{1.18}$ mass-to-light ratio adopted by 
Laor (2001), our best-fitting $M_{bh}-L_{bulge}$ relation is
equivalent to $M_{bh} \propto M_{bulge}^{1.06\pm0.06}$,
again, completely consistent with a linear scaling. 
\begin{figure}[h]
\begin{center}
\includegraphics[width=.5\textwidth,angle=270]{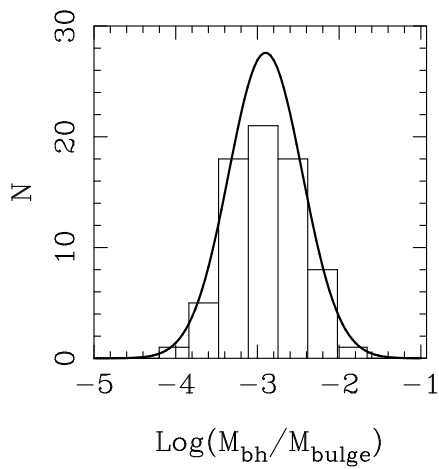}
\end{center}
\caption[]{Histogram of the ratio of black-hole mass to bulge mass for
the 72-object AGN sample. Over-plotted for comparison is a gaussian
with $\langle \log (M_{bh}/M_{bulge}) \rangle =-2.90$ and standard deviation
0.45 (see text for discussion)}
\label{fig2}
\end{figure}
\subsection{The normalization of the black hole - bulge mass relation}
Having established that the $M_{bh}-M_{bulge}$ relation is consistent 
with being linear, we now assume perfect linearity in order to
establish the normalization of the $M_{bh}-M_{bulge}$ relation. With the 
mass-to-light ratio adopted here, a linear scaling corresponds to 
enforcing a slope of $-0.524$ in the $M_{bh}$ vs. $M_{R}$ relation. 
Under this restriction the best-fitting relation has a normalization of
$M_{bh}=0.0012M_{bulge}$, and can clearly be seen to be an 
excellent representation of the data (Fig 1). It is noteworthy that the 
normalization of $M_{bh}=0.0012M_{bulge}$ is identical to that determined by
Merritt \& Ferrarese (2001b) from their velocity dispersion study of
the 32 inactive galaxies in the Magorrian et al. sample. 

The closeness of the
 agreement between the $M_{bh}/M_{bulge}$ ratios determined here with
 those determined by Merritt \& Ferrarese is highlighted by Fig
 \ref{fig2}, which shows a histogram of the $M_{bh}/M_{bulge}$
 distribution for our 72-object AGN sample. The AGN $M_{bh}/M_{bulge}$ 
distribution has $\langle \log (M_{bh}/M_{bulge})\rangle =-2.87
 \pm{0.06}$ with a standard deviation of $\sigma=0.47$. This is in 
remarkably good agreement with the Merritt \& Ferrarese results, which were 
$\langle \log (M_{bh}/M_{bulge}) \rangle =-2.90$ and
 $\sigma=0.45$. Finally, we note that the normalization of 
$M_{bh}=0.0012M_{bulge}$ agrees very well with the predictions of 
recent models of coupled bulge/black hole formation at high redshift 
(Archibald et al. 2001).

\section{Bulge luminosity versus stellar velocity dispersion}
\begin{figure}[h]
\begin{center}
\includegraphics[width=1.\textwidth]{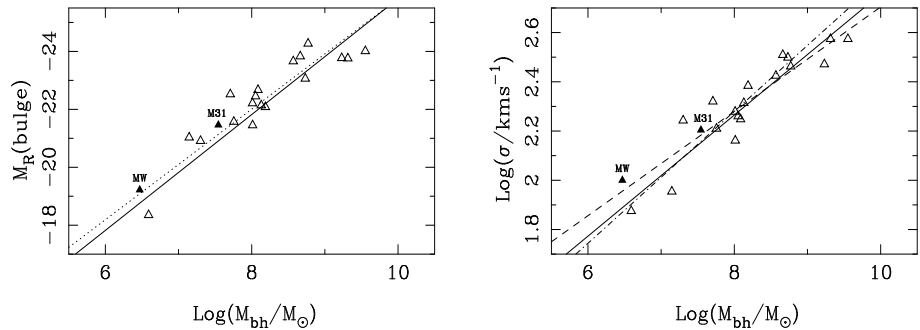}
\end{center}
\caption[]{Left-hand panel shows absolute $R-$band bulge magnitude
versus dynamical black-hole mass estimate for our inactive
galaxy sample. The solid line is the best-fitting relation 
($M_{bh} \propto M_{bulge}^{0.95\pm0.09}$) and the dotted
line is the best-fitting linear relation ($M_{bh}=0.0012M_{bulge}$). 
The right-hand panel is the same with bulge luminosity replaced by stellar velocity
dispersion. The solid line is the best-fit ($M_{bh}\propto \sigma^{4.09}$)
, the dashed line is the
Merritt \& Ferrarese (2001a) relation ($M_{bh}\propto \sigma^{4.72}$),  and
the dot-dashed line is the Gebhardt et al. (2000) relation 
($M_{bh}\propto \sigma^{3.75}$). The location of the 
Milky Way and M31 are indicated for the interest of the reader,
although neither were included in the analysis.}
\label{fig3}
\end{figure}
The quality of the fit to the inactive galaxy sample is 
illustrated by the left-hand
panel of Fig \ref{fig3}, which shows the $M_{bh}-L_{bulge}$ relation
for the inactive galaxy sample alone. Of particular interest is
the scatter around this best-fit relation, given that it has been widely
reported in the literature (eg. Merritt \& Ferrarese 2001a, Kormendy \&
Gebhardt 2001) that the scatter around the $M_{bh}-L_{bulge}$ relation 
is significantly greater than that around the $M_{bh}-\sigma$
relation. However, in contrast, we find that the scatter 
around the $M_{bh}-L_{bulge}$ relation for our sample of 
nearby inactive galaxies, which excludes non E-type
morphologies, is only $0.33$ dex, in excellent agreement with 
the scatter around the $M_{bh}-\sigma$ 
relation (Merritt \& Ferrarese 2001a).

To test this result further, in the right-hand panel of 
Fig \ref{fig3}, we investigate the $M_{bh}-\sigma$ relation for 
our nearby inactive galaxy sample. The scatter around the 
best-fit relation ($M_{bh} \propto \sigma^{4.09}$) is 0.30 dex, 
leading us to the conclusion that the intrinsic 
scatter around the $M_{bh}-L_{bulge}$ relation for {\it elliptical} galaxies
is comparable to that in the M$_{bh}-\sigma$ relation.

\section{Conclusions}
The main conclusions of this study can be summarized as follows:
\begin{itemize}
\item{The best-fitting  $M_{bh}-L_{bulge}$ relation to the combined 
sample of 72 AGN and 18 nearby inactive elliptical galaxies is found to be 
consistent with a linear scaling between black hole and bulge mass
($M_{bh}\propto M_{bulge}^{0.95\pm0.05}$), and to have much lower
scatter than previously reported ($\Delta \log M_{bh}=0.39$ dex).}
\item{The best-fitting normalization of the $M_{bh}-M_{bulge}$ relation is 
found to be $M_{bh}=0.0012M_{bulge}$, in excellent agreement with 
recent stellar velocity dispersion studies.}
\item{In contrast to previous reports it is found that the
scatter around the $M_{bh}-L_{bulge}$ and $M_{bh}-\sigma$ relations
for nearby inactive elliptical galaxies are comparable, at
only $\sim 0.3$ dex. }
\end{itemize}

%

\end{document}